# Hybrid graphene plasmonic waveguide modulators

D. Ansell[1], B. D. Thackray[1], D. E. Aznakayeva[1], P. Thomas[1], G. H. Auton[1], O. P. Marshall[1], F. J. Rodriguez[1], I. P. Radko[2], Z. Han[2], S. I. Bozhevolnyi[2] and A. N. Grigorenko[1*]

[1]*School of Physics and Astronomy, University of Manchester, Manchester, M13 9PL, UK*

[2]*Institute of Technology and Innovation (ITI), University of Southern Denmark, Niels Bohrs Allé 1, DK-5230 Odense M, Denmark*

The unique optical and electronic properties of graphene allow one to realize active optical devices. While several types of graphene-based photonic modulators have already been demonstrated, the potential of combining the versatility of graphene with subwavelength field confinement of plasmonic/metallic structures is not fully realized. Here we report fabrication and study of hybrid graphene-plasmonic modulators. We consider several types of modulators and identify the most promising one for light modulation at telecom and near-infrared. Our proof-of-concept results pave the way towards on-chip realization of efficient graphene-based active plasmonic waveguide devices for optical communications.

* correspondence and requests for materials should be addressed to A.N.G.
(email: sasha@manchester.ac.uk).

## Introduction

Graphene holds a great potential to provide efficient graphene-based photodetectors[1-4], dedicated sensors[5-7] and various optoelectronic devices[8-10]. Tunability of graphene conductivity by gating[11-14] should allow one in principle to realize compact optical modulators. While several types of graphene-based photonic modulators have already been demonstrated[15-19], the combination of plasmonic waveguides with graphene for the task of light modulation remains elusive. Hybrid graphene modulators[15-19] promise unrivalled speed, low driving voltage, low power consumption, and small physical footprints. Such modulators will be welcomed by telecommunications and optoelectronics.

To increase graphene interaction with light, one can use conventional metal plasmonics[3, 9, 20-24]. Plasmonic waveguides could provide smaller volumes of propagating modes and local field enhancement which would result in higher modulation depth. There are two main problems in combining graphene and plasmonic waveguides in order to achieve optical modulation by gating: i) the presence of metal layers complicates the task of gating with the spacer dielectrics being often affected by both high electric fields[25] and light[26], and ii) it is difficult to realize plasmonic waveguide modes with large in-plane field components, since fields of surface plasmon-polaritons (SPPs) supported by metal-dielectric interfaces are predominantly transverse in dielectrics[27] (and the perpendicular electric fields do not excite currents in recumbent graphene). While the latter feature provides a useful possibility to safeguard metal plasmonics with graphene without degrading plasmon characteristics[22], it makes the endeavour of designing hybrid graphene plasmonic modulators rather challenging – yet not hopeless.

Here we discuss two types of graphene modulators based on plasmonic waveguides[27] and collective plasmons[24]. We also describe unusual properties of hafnium oxide as a dielectric separator for graphene based active optical devices.

## Hybrid graphene plasmonic waveguide modulator (HGPWM)

The most straightforward HGPWM geometry modulated by the Pauli blocking effect (Fig. 1**a**) is based on the classical SPP configuration and shown in Fig. 1**b**, top inset, where the gold strip (yellow colour), which supports the SPP propagation and serves as a backgate, is covered by a hexagonal boron-nitride (hBN) flake (purple colour), acting as a dielectric spacer and atomically smooth substrate for graphene, and a graphene flake (black colour). However, the SPP mode in such a waveguide configuration is, away from the surface plasmon resonance in the visible, only weakly bound to the metal surface and features primarily transverse electromagnetic fields, which do not excite currents in graphene, while in-plane fields (which do interact with the graphene in-plane conductivity) are negligibly small in the infra-red[28]. Hence, the classical HGPWM configuration, producing only very weak graphene-related absorption and promising thereby only very weak modulation by gating, can hardly be used in practice. Attempting to enhance in-plane field components in the graphene layer, we introduced a nanostructured (corrugated) part of the plasmonic waveguide so as to produce longitudinal near fields generated by the SPP mode propagating along the corrugated part of waveguide (see middle inset of Fig. 1**b**). It is however clear that the expected enhancement of in-plane field components is quite limited as the metal surface corrugation has to be shallow in order to not introduce significant additional propagation losses by scattering. Finally, the wedge SPP mode supported by the edge of planar section of the waveguide[29, 30] turned out to be useful for the tasks of light modulation (see bottom inset of Fig. 1**b** and Fig. 1**c**).

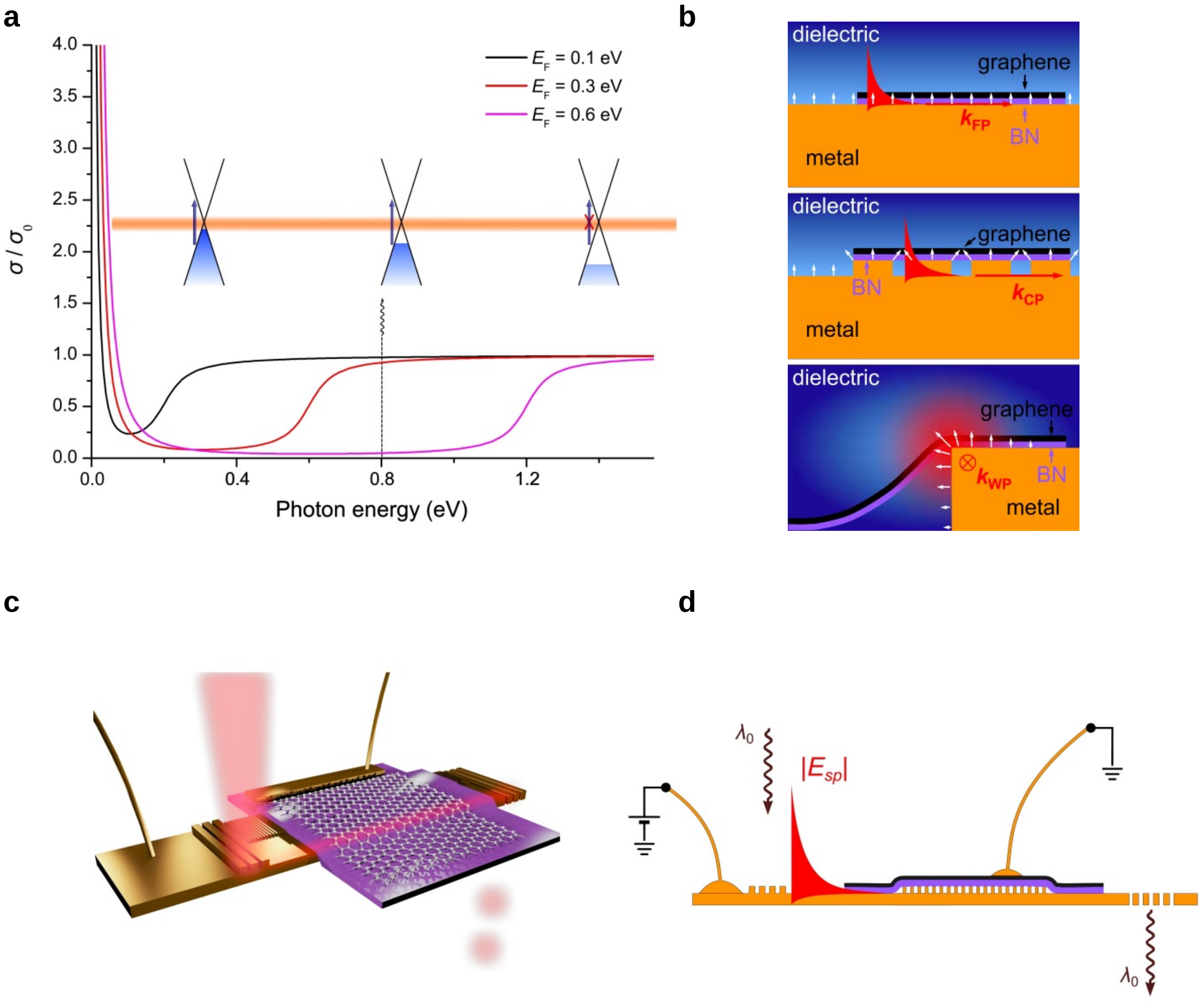

Figure 1. Principle of hybrid graphene plasmonic waveguide modulators. a, Optical Pauli blocking. b, Sketches of three types of plasmonic modes – flat, corrugated, and wedge plasmons. White arrows indicate approximate direction of electric field. c, 3D rendering of the experiment with the wedge plasmon mode. d, The schematic of experiment where non-transparent grating couples light into plasmon modes which can be affected by gated graphene, black layer, placed on the top of dielectric spacer (a flake of boron nitride, purple layer) and then be decoupled into light through the transparent grating.

The wedge mode, in addition to enhanced in-graphene-plane fields near the edge of the strip that should result in higher modulation depth induced by graphene gating, has superior field confinement characteristics, which is essential when considering potential applications to surface plasmon circuitry. Figure 1**d** provides a general outline of modulation experiments: a non-transparent gold grating couples light into a plasmon-propagating mode which can be affected by gated graphene placed on the top of dielectric spacer and then running plasmons are decoupled into light through the transparent grating. Such configuration allows one to decrease the crosstalk between input and output light[27].

All three studied plasmon-polariton modes – flat plasmons (FP), corrugated plasmons (CP) and wedge plasmons (WP) – can be excited by moving the incident light beam to different parts of the coupler, Fig. 2**a**. An optical micrograph of one of our devices studied in this work is shown in Fig. 2**b** along with outlines demonstrating positions of hBN and graphene flakes. We have checked operation of plasmonic waveguides in both transmission and leakage radiation[31, 32] modes. Leakage radiation detection of plasmonic propagating modes for wedge and flat plasmons are shown in Fig. 2**c**. Figure 2**c** confirms that the plasmonic modes were successfully excited and propagated along the waveguide. For completeness, Fig. 2**d** provides a SEM micrograph of an area marked in Fig. 2**b** by the blue dashed box where the semi-transparent decoupler and a part of the nanostructured area of the waveguide are shown.

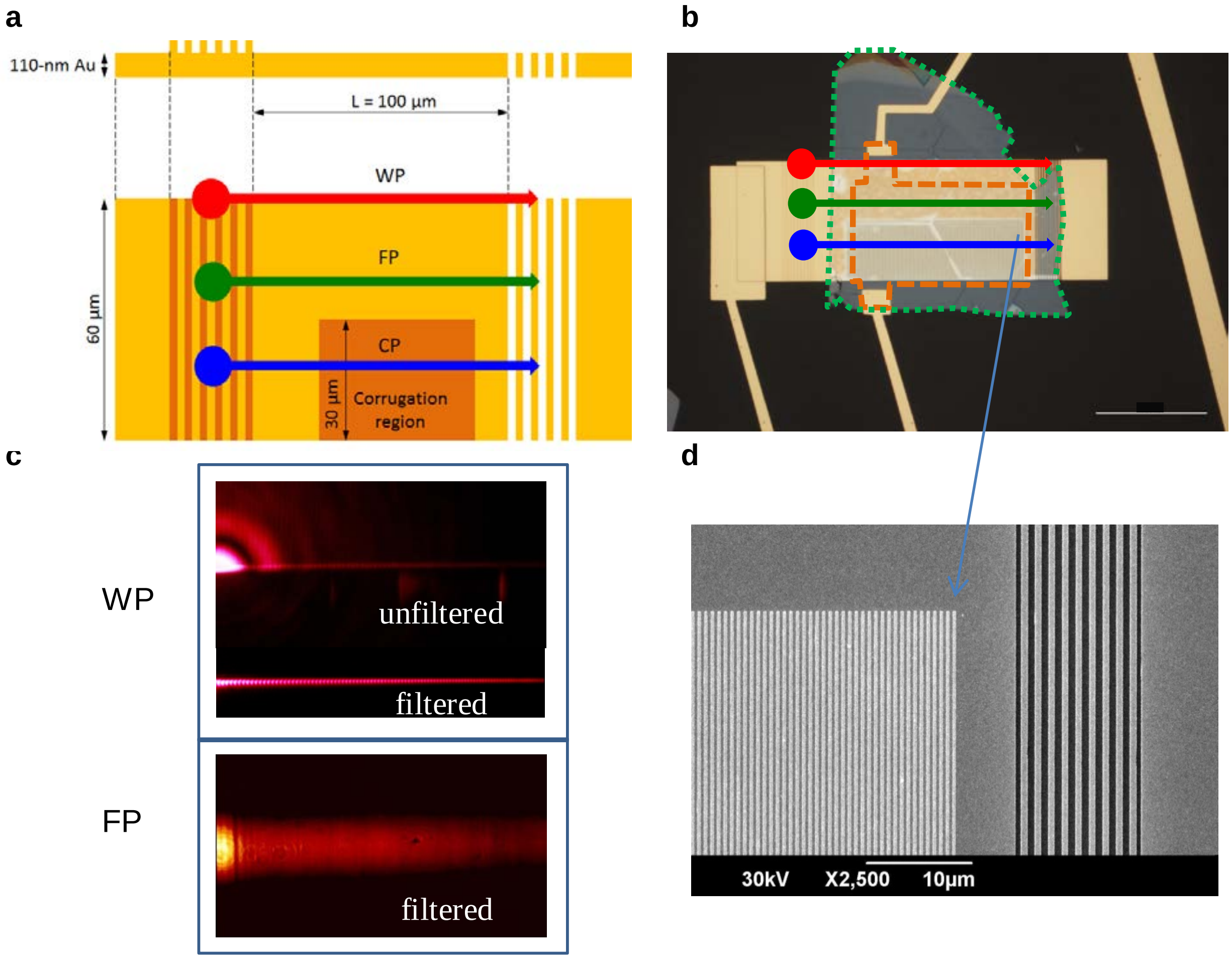

Figure 2. Plasmon modes of hybrid graphene plasmonic waveguide modulators. a, The schematics of a studied plasmonic waveguide. The red, green and blue arrows represent wedge plasmon mode (WP), flat plasmon mode (FP) and corrugated plasmon mode (CP), respectively. **b,** The optical micrograph of a typical hybrid graphene plasmonic modulator studied in this work. The red, green and blue arrows represent wedge, flat and corrugated plasmon modes, respectively. An area enclosed by green dotted line represents hBN. An area enclosed by dotted brown line represents graphene. The scale bar is 50 µm. **c,** Leakage radiation detection of wedge, upper panel, and flat, lower panel, plasmon propagating modes. The wedge mode is given in both raw and Fourier filtered images. **d,** A scanning electron micrograph of an area shown in **b** by the dotted box that shows corrugated waveguide and the semi-transparent decoupling grating.

The plasmonic waveguides were excited by telecom laser providing ~ 3 mW of power at wavelength $\lambda$ = 1.5 µm. We measured the dynamic response of our modulators by applying an offset square-wave voltage to the back gate with peak-to-peak amplitude $V_g^{pp}$ and dc component $V_g^{dc}$. Figure 3 shows the modulation-depth characteristics both as a function of $V_g^{pp}$ and $V_g^{dc}$. Comparing the modulation depth of FP and CP modes (Fig. 3**a**), one can see that the modulation effect is substantially stronger for CP: the CP mode gives around an eightfold increase in modulation depth compared to FP (for large $|V_g^{dc} - V_{CNP}|$). Here $V_g^{pp}$ is set to 7.6 V and 6 V for the measurements of the FP and CP mode modulation depths, respectively. For both FP and CP we see an approximately symmetrical increase in modulation depth with $V_g^{dc}$, which we attribute to being the positions of the CNP (for the FP mode this is $V_{CNP} \approx 2.7$ V; for the CP mode this is $V_{CNP} \approx 0.9$ V). For other samples the modulation for FP modes compared to CP modes was even less pronounced. A drastic improvement in modulation depth is expected from the increased interaction of graphene with electric field of the CP mode. Indeed, the longitudinal component of electric field of FP is rather weak, whereas the presence of corrugations creates strong local longitudinal fields near ridges (see the sketch of the mode on Fig. 1**b** and, e.g., Ref.[33]).

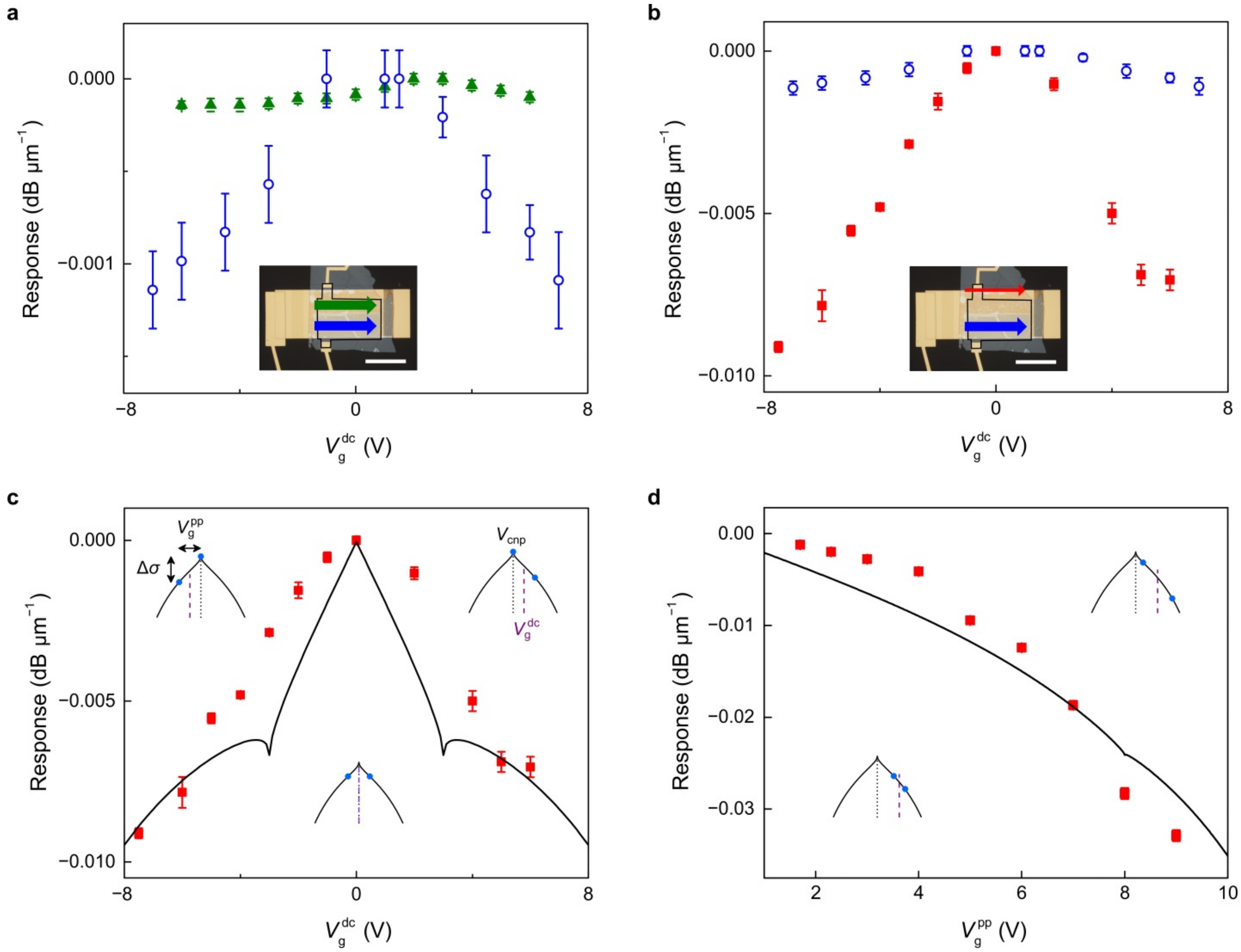

Figure 3. Operation of hybrid graphene plasmonic waveguide modulators. a, Modulated AC transmission of the waveguide expressed in dB/µm as a function of gating voltage for the flat, green data points, and the corrugated, blue data points, plasmon modes. The filled and empty data points represent two different devices. The inset shows the position where the plasmonic modes were measured. **b,** Modulated AC transmission of the waveguide expressed in dB/µm as a function of gating voltage for the wedge, red data points, and the corrugated, blue data points, plasmon modes. The inset shows the position where the plasmonic modes were measured. **c,** Theoretical fit to the experimental data for the wedge mode modulation as a function of gating. **d,** Modulated AC transmission of the waveguide expressed in dB/µm as a function of AC peak-to-peak amplitude for the wedge, red data points, and the corrugated, blue data points, plasmon modes.

The wedge plasmon mode provided us with the greatest modulation depths, achieved when probing the top edge of the waveguide, see inset Fig. 3**b**. Similarly to FP and CP, the modulation depth of WP increases symmetrically from an offset gate voltage $V_g^{\mathrm{dc}}$, which in this case is $\approx 0$ V and is close to $V_{\mathrm{CNP}}$ observed in dc resistivity measurements. For $V_g^{\mathrm{dc}} = 6$ V and $V_g^{\mathrm{pp}} = 9$V we were able to achieve a modulation depth of $3.3 \times 10^{-2}$ dB $\mu\mathrm{m}^{-1}$, around 30 and 230 times larger than the best CP and FP modulation depths with similar set parameters, respectively, Fig. 3**d**. It is important to note that the measured modulation depth for WP is underestimated, because, due to the sample geometry, it was not possible to excite solely the WP mode – the FP mode was always excited on the adjacent flat region. As a result, the detected modulated WP signal had a contribution from FP, which was modulated by a substantially lower degree. A more than an order-of-magnitude increase of the modulation values of WP over CP (Figs. 3**b** and 3**d**) can be again attributed to an increased in-plane (in this case, lateral) electric-field component of the WP mode, which is essential in excitation of currents in graphene (see the sketch of the mode on Fig. 1**b**). The largest modulation depth observed for WP mode was about 8.7% for 12 μm modulation length. Detailed discussion of the operation of HGPWM is in Ref. 27.

## Graphene modulator based on diffractive coupled plasmons

Plasmon resonances in metallic nanostructures have attracted a lot of interest in recent years for their promising applications spanning many fields, from negative index metamaterials and perfect lensing[34] to extremely sensitive bio-sensing[35, 36] . For key applications, e.g. sensing and active plasmonics, it is crucial to have the narrowest plasmon resonances possible. Surface plasmon polaritons propagating in a continuous film generally provide a resonance quality factor at the level of Q < 20 [37], while localized plasmon resonances observed in isolated nanoparticles tend to be even broader, limiting their usefulness for the tasks of bio-detection and light modulation.

The spectral width of the resonance peak can be narrowed by coupling resonance modes in regular nanostructure arrays and thus providing ultra-narrow collective, diffraction coupled plasmon resonances[38-41], also known as geometric resonances. If arrays are fabricated so that the Rayleigh diffraction anomalies of the array[42] (where light is diffracted parallel to the plane of the array as a diffraction mode crosses from the air into the substrate) and the localized surface plasmon resonance modes of the structures occur at similar wavelengths, then light that would otherwise be scattered to the far-field can be recaptured as electron oscillations in the neighboring nanostructures in phase with plasmon excitation induced by the incident light. By using the right combination of nanostructure size, shape and array period, one can achieve ultra-narrow and deep collective plasmon resonances at the desired wavelength that normally improve with increasing array size and/or spatial coherence of the beam.

Recently, we achieved a significant improvement in the quality factor of the collective resonances designed for telecom wavelengths[24]. We have measured some of the highest recorded values of quality factor for collective resonances in diffraction coupled arrays of plasmonic nanostructures, registering Q ≅ 300 at wavelengths of around 1.5 μm. This improvement was achieved by adding a continuous gold layer beneath a gold nanostripe array and utilizing a large angle of incidence (around 80°) which generates large image dipoles in the gold sub-layer and mimics an index-matched environment. The typical high quality plasmonic resonance based on diffraction coupled plasmons measured in our samples is shown in Fig. 4. We estimate a quality factor $Q \cong 300$ at a wavelength of $\lambda$ = 1515 nm at an angle of incidence of $\theta$ = 80° while the width of the resonance is around 5nm, see below. The sample has lines 420 nm wide with a period $a$ = 1530 nm and were fabricated on a 65 nm gold film. The resonance position corresponds very well to the expected position of the Rayleigh cut-off wavelength[40, 42] for air at $\lambda_R = \frac{a}{m}[1 + \sin\theta]$, for $m$ = 2. Peaks corresponding to the $m$ = 3, 4 and 5 resonance modes are also present, Fig. 4. It is important to stress that the control sets of stripes with the same sizes and periodicities fabricated on a bare glass substrate did not show narrow resonances at the same conditions, see an example in Fig. 4. Therefore, the gold sublayer beneath the stripe array allowed us to suppress the negative effect of the substrate on collective resonances in an asymmetric refractive index environment.

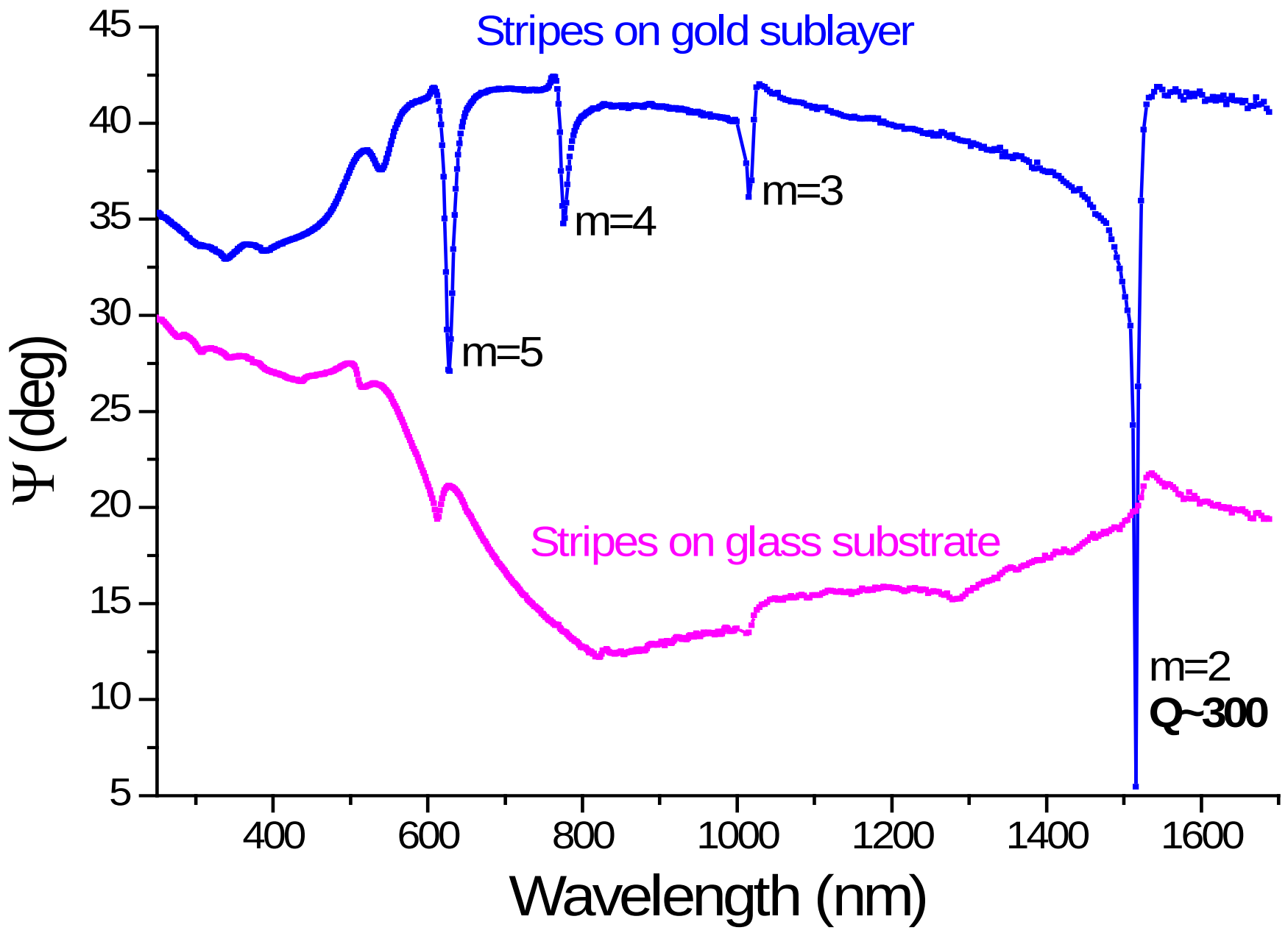

Figure 4. Diffraction coupled plasmon resonance measured in our nanostripe arrays at the incidence angle of 80°. Stripes width is 420 nm wide, stripe period is a = 1530 nm. The blue curve shows the collective resonances of stripes fabricated on a gold layer; the magenta curve gives the ellipsometric reflection for the control stripes made on bare glass substrate.

One promising application of diffraction coupled resonances lies in the field of active plasmonics: sharp plasmonic features could allow one to realise strong light modulation. There has been a lot of interest in the development of plasmonic devices which combine noble metal metamaterials with two-dimensional materials such as graphene[43]. One of the goals of graphene-plasmonic hybrid devices is the modulation of light by graphene gating[23, 44]. The schematic and the operation of the studied hybrid devices are shown in Fig. 5. To achieve graphene gating we have added boron-nitride/graphene layers to the plasmonic structure which resulted in a moderate reduction of both the resonance quality and its depth.

We applied a gate voltage up to +/– 150 V to our graphene/BN/nanostripe array device (pictured in Fig. 5a) and measured the change in reflectivity of *p*-polarised light ($R_p$). Figure 5b shows that applying a gate voltage redshifts the $m = 2$ resonance by approximately 10 nm at an angle of incidence of 70° (this angle is moderately different from the optimal due to the presence of BN and the finite size of the graphene flake). Figure 5b shows that the redshift of the resonance induced by gating was large enough to change the measured reflection $R_p$ value on the right-hand side of the resonance at $\lambda = 1535$ nm by 20%. (At the same time, it is worth noting that gating does not change the depth of the resonance considerably.) The FDTD modelling of the response of our structure is also shown in Fig. 5b as dashed lines.

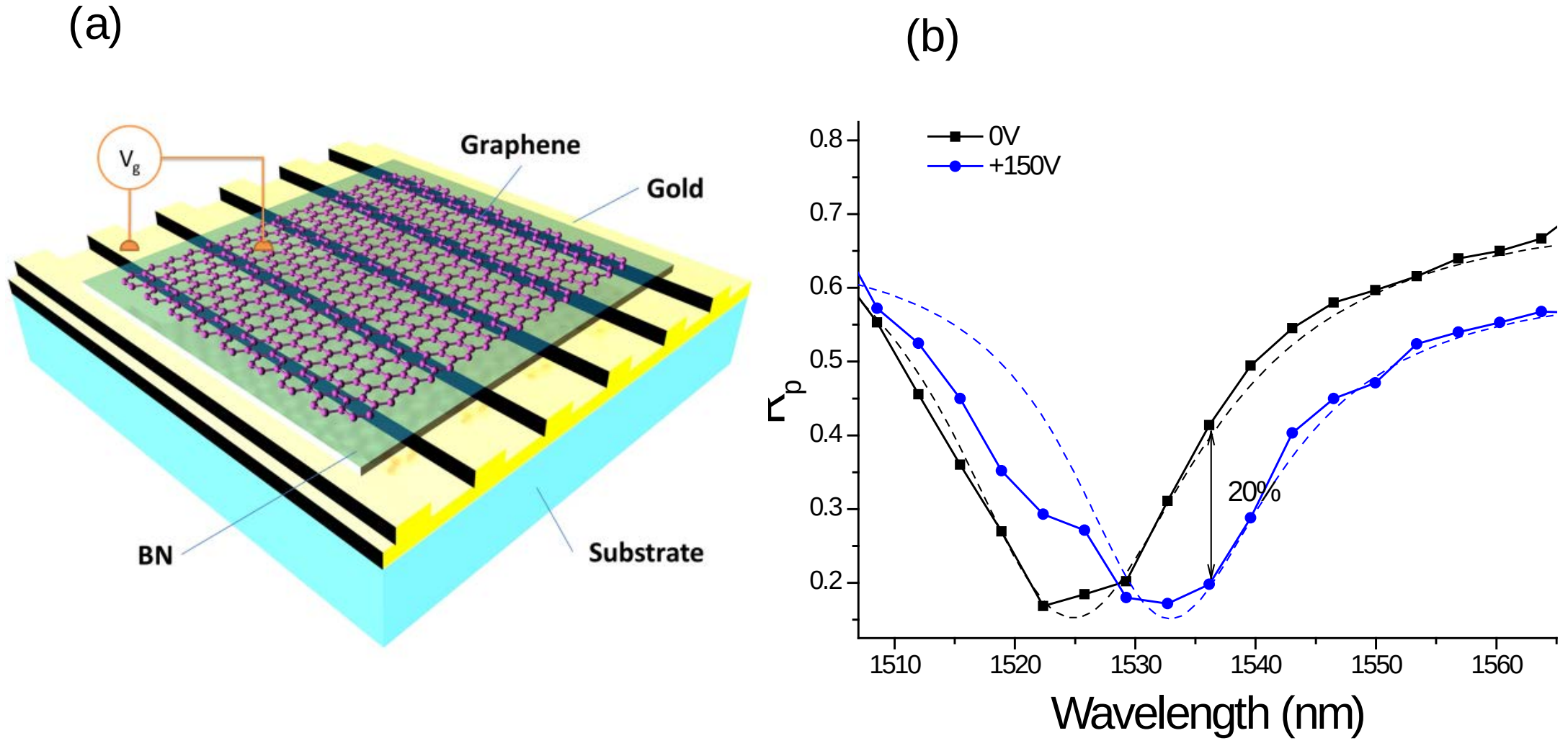

Figure 5. Electro-optical modulation of plasmon resonance using graphene gating. (a) Schematics of a hybrid graphene plasmonic optical modulator. (b) Change of p-polarized reflection $R_p$ at an incident angle of 70° for the $m$=2 plasmon resonance mode due to graphene gating. The dashed lines provide the reflection of the structure calculated with the help of FDTD simulations.

## Low-voltage open-air graphene modulators.

The main disadvantage of the modulators described above is relatively large voltages needed to induce large modulation of the output light intensity (~10V in the case of the HGPWM and ~100V in the case of the modulator based on the collective resonances). This is connected to the fact that hBN thickness needs to be relatively large in order to avoid electrical breakdown, photoconductivity and electrical hysteresis. It would be ideal to reduce the modulating voltage to ~1V in order to be competitive with, e.g., liquid crystal devices. To explore this possibility, we made a set of simple capacitance type devices with different dielectrics as graphene separator from the bottom contact made of copper. We found that the modulator produced by heterostructures made of metal (copper), high-k-dielectric ($HfO_2$) and graphene can indeed be modulated at the voltages of around ~1V. It also provides small modulation volume, has low power consumption (< 1μW) and shows modulation coefficient K = 3.3 % at λ=1.5 μm, , K = 2 % at the λ=1.3 μm, K = 1.5 % at λ=1 μm, K = 0.6% at λ=0.9 μm using a single layer of graphene. These devices are CMOS compatible and could find a wide range of applications.

Figure 6 shows schematics of the fabricated samples. The bottom layer of the modulator structure was copper with the thickness of 35 and 70 nm. The function of the bottom layer is to serve as the reflective mirror and back gate electrode. To eliminate copper oxidation and its properties degradation the quarter wavelength high-k-dielectric layer of $HfO_2$ has been chosen to be deposited afterwards on the top of copper layer. The graphene monolayer was placed on the top of $HfO_2$ dielectric separator. The thickness of hafnium oxide layer ($\lambda/4n$, where n is the refractive index of hafnium oxide) was chosen in such a way that interference superposition between incident electromagnetic wave and reflected wave from the bottom copper mirror, would lead to the formation of standing wave with the maximum amplitude of electric field acting on the graphene layer.

(a) (b)

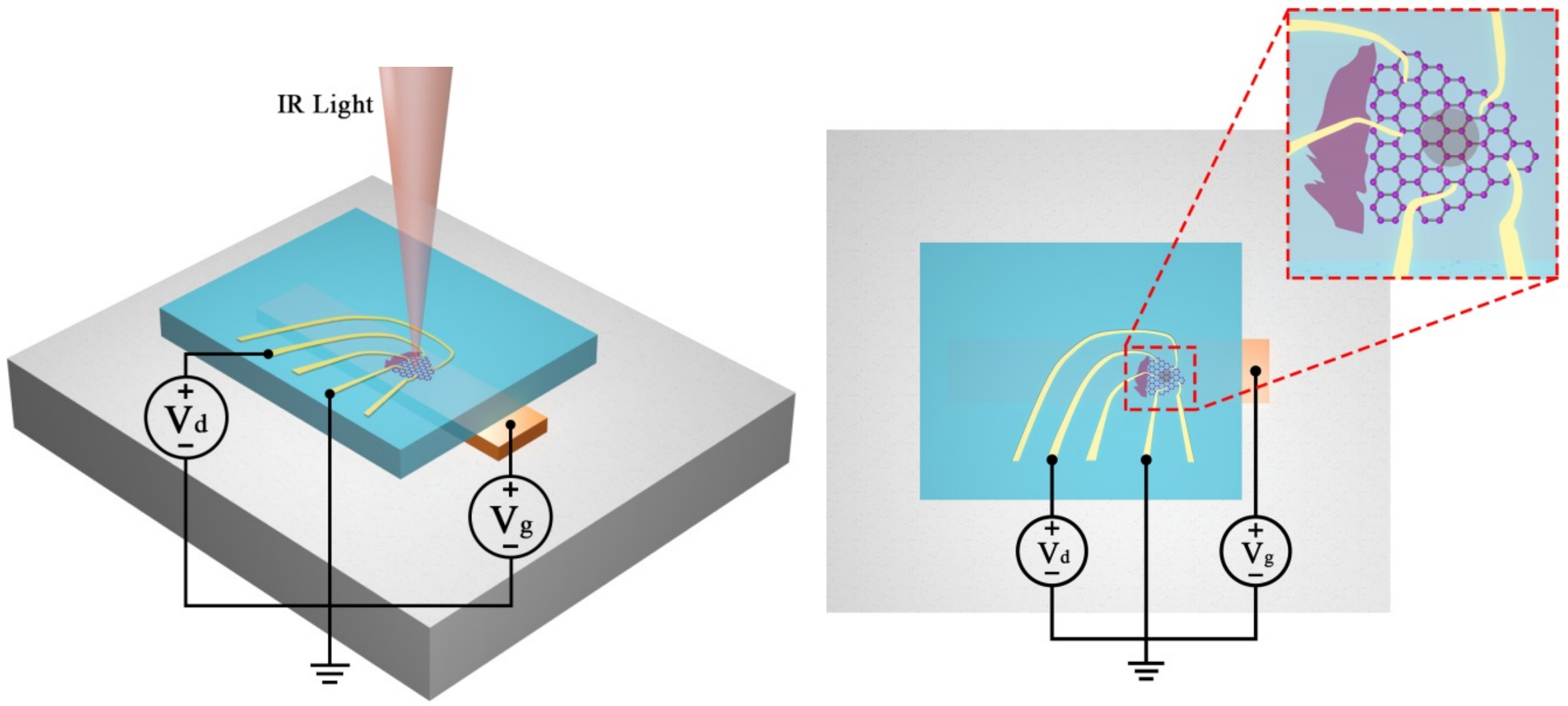

Figure 6. Graphene based electro- optical modulator. (a) 3D rendering, (b) A view from the top. The modulator structure consists of quartz, reflective copper mirror, sub wavelength high-k $HfO_2$ dielectric layer and defect free high quality mechanically exfoliated graphene monolayer.

The optical properties of the fabricated devices were measured using Brucker FTIR spectrometer working in the reflection mode. Figure 7(a) gives the relative reflection of the structures measured with respect to reflection from a gold mirror. Figure 7(b)-(c) shows the reflection spectra of fabricated devices at several gating voltages (measured relative to the response at zero voltage). The observed value of relative electro-optical modulation coefficient K = 3.3 % achieved at the telecommunication wavelength λ = 1.5 μm, K = 2 % at the telecommunication wavelength λ = 1.3 μm, K = 1.5 % at the near infrared wavelength λ=1.3 μm at low gating voltage of $V_g$= - 2 V. More importantly, the modulation extends to the wavelengths λ = 0.9 μm with modulation coefficient of K = 0.6% at gating voltage of 3 V. This implies that hafnium oxide can be used as the dielectric separator for graphene based optical modulators working at low voltages. The detailed discussion of these modulators along with relevant physics which allows their operation will be described elsewhere.

To conclude, we have shown that i) the use of wedge SPP waveguide configuration for light modulation by graphene gating paves thereby the way towards practical realization of very compact and efficient, potentially ultrafast[1] and broadband[11] hybrid graphene-plasmonic optical devices for wide range of applications in optoelectronics and telecommunications; ii) collective plasmon resonances can be used to achieve large modulation depth even in case of a single graphene layer incorporated into a hybrid graphene plasmonic device; iii) hafnium oxide can provide a viable alternative to hBN as a dielectric separator for low voltage graphene based modulators.

(a)

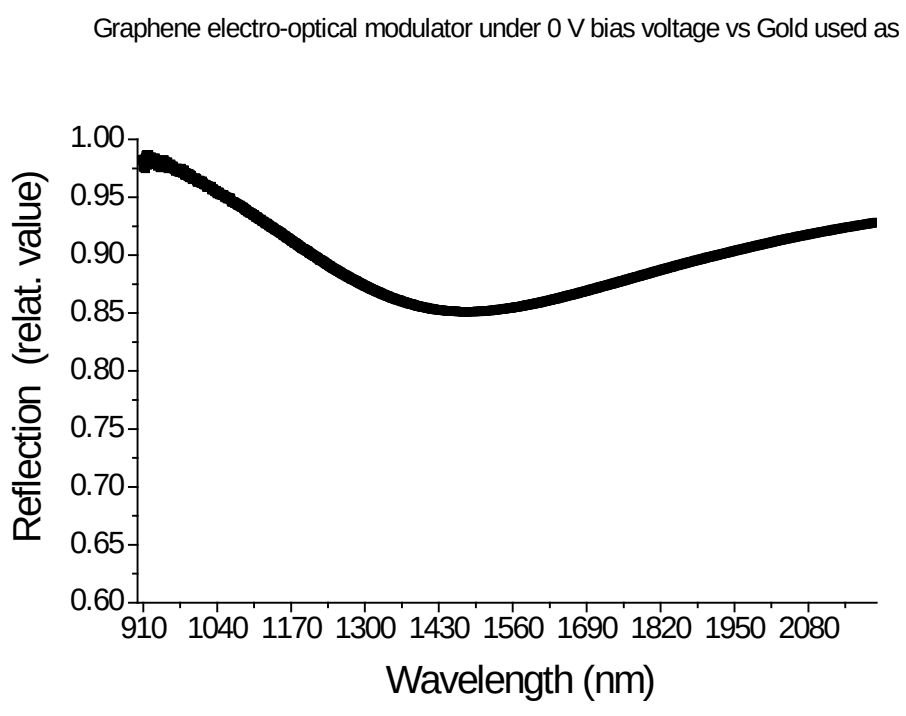

(b)

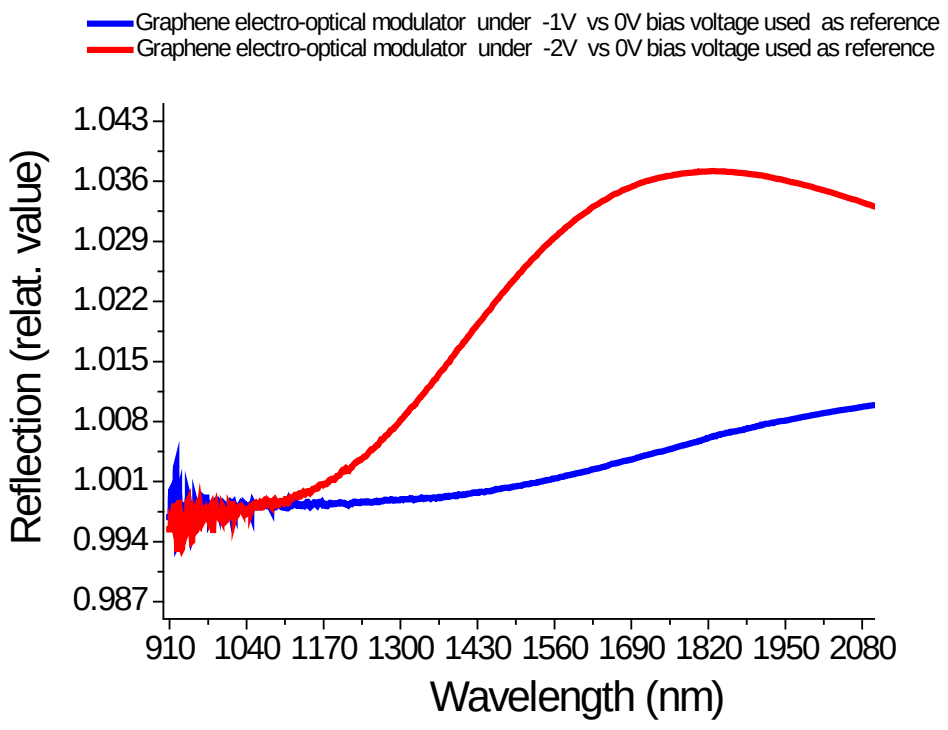

(c)

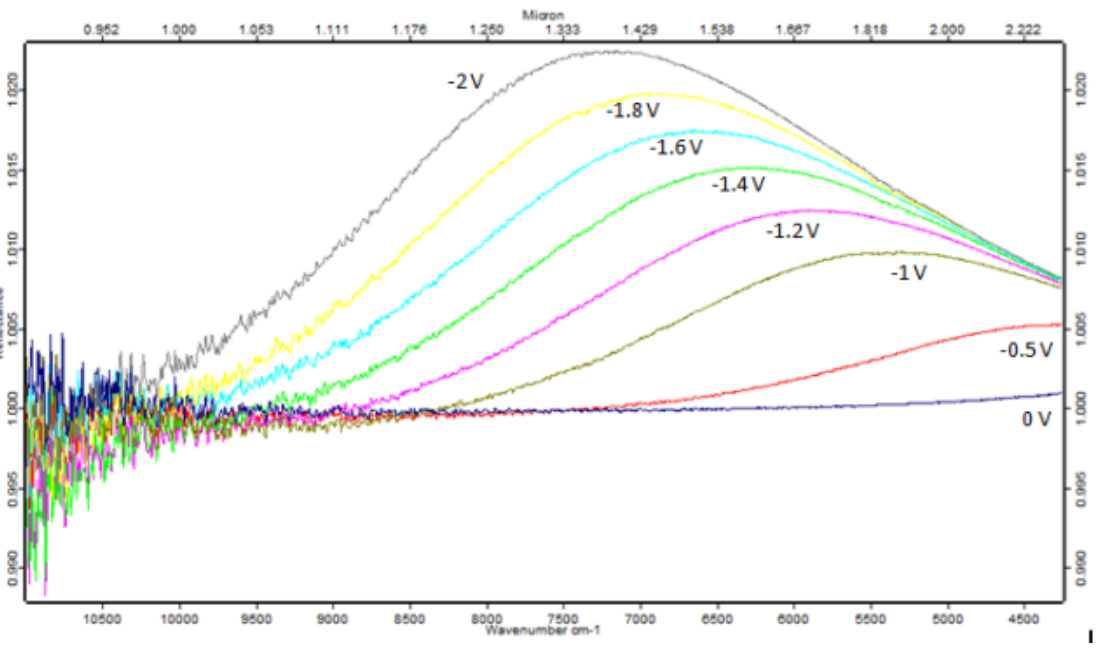

(d)

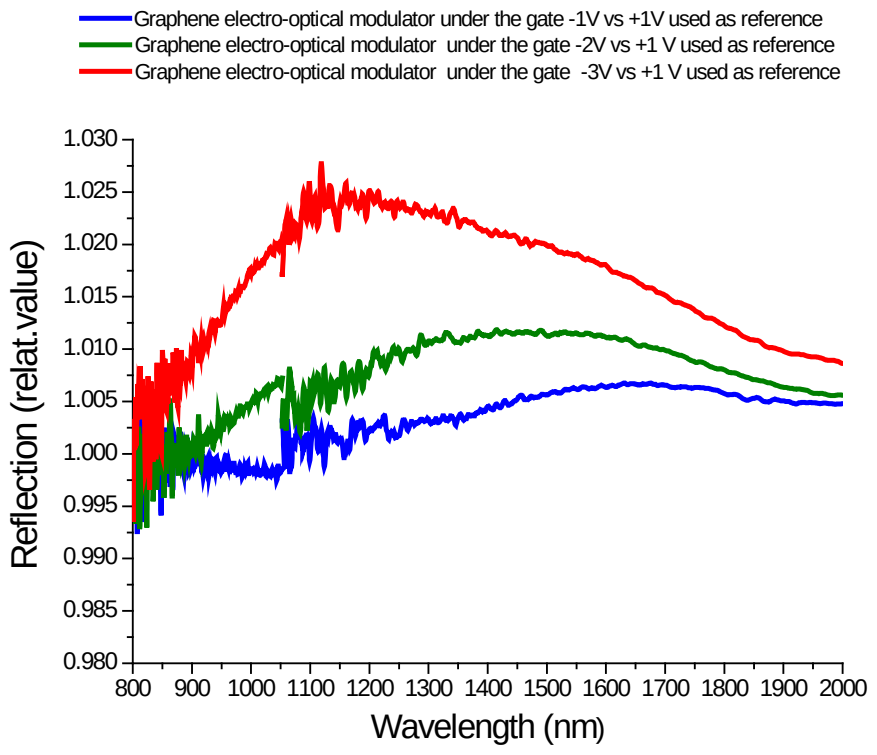

Figure 7. Reflection spectra of graphene based modulators. (a) reflection relative to gold for a modulator operating at λ=1.5 μm at zero gating voltage (b) modulation of the reflection due to applied gating voltage of -1 V (blue curve) and - 2 V (red curve) relative to zero voltage reflection. The value of the modulation coefficient K = 3.3% at λ=1.5 μm. (c) reflection spectra of the modulator operating at λ=1.3 μm at 0 V gate voltage (blue curve), -1 V (yellow curve),- 2 V (black curve) applied to the device. The value of modulation coefficient K = 2% at λ=1.3 μm. (d) reflection spectra of the modulator operating at λ=1 μm at 0 V DC gate voltage (blue curve), -1 V (green curve),- 2 V (red curve) applied to the device. The value of modulation coefficient K = 1.5% at λ=1 μm.

**Acknowledgements**

ANG, DA and FJR are grateful to the support by SAIT GRO Program, EPSRC grant EP/K011022/1, Bluestone Global Inc. Grant, E.C. Graphene Flagship grant (contract no. CNECT-ICT-604391). IPR, ZH and SIB acknowledge financial support for this work from the Danish Council for Independent Research (the FTP project ANAP, Contract No. 09-072949) and from the European Research Council, Grant No. 341054 (PLAQNAP).